\documentclass[runningheads]{llncs}

\usepackage{comment}
\usepackage{amssymb} 
\usepackage{color}
\usepackage[dvipsnames]{xcolor}
\usepackage{acronym, amsmath, booktabs, graphicx, makecell, multirow, subcaption, tabularx, wrapfig, xcolor}
\acrodef{ADC}{Analog-to-Digital Converter}
\acrodef{CCD}{Charged Coupled Device}
\acrodef{CMOS}{Complementary Metal Oxide Semiconductor}
\acrodef{CFA}{Color Filter Array}
\acrodef{HDR}{High Dynamic Range}
\acrodef{MLE}{Maximum Likelihood Estimation}
\acrodef{CDF}{Cumulative Distribution Function}
\acrodef{PDF}{Probability Density Function}
\acrodef{PMF}{Probability Mass Function}
\acrodef{ISP}{Image Signal Processing}
\acrodef{ND}{Neutral Density}
\acrodef{MVUE}{Minimum Variance Unbiased Estimator}
\acrodef{CRF}{Camera Response Function}
\acrodef{SNR}{signal-to-noise ratio}
\acrodef{CRLB}{Cram\'er–Rao Lower Bound}
\acrodef{EM}{Expectation Maximization}
\acrodef{MC}{Monte Carlo}
\acrodef{IQA}{Image Quality Assessment}

\newcolumntype{Y}{>{\centering\arraybackslash}X}

\definecolor{Gold}{rgb}{0.99, 0.76, 0.0}
\definecolor{Silver}{rgb}{0.79, 0.75, 0.73}
\definecolor{Bronze}{rgb}{0.8, 0.5, 0.2}
\newcommand{\first}[1]{\colorbox{Gold}{#1}}
\newcommand{\second}[1]{\colorbox{Silver}{#1}}
\newcommand{\third}[1]{\colorbox{Bronze}{#1}}

\begin{document}
\pagestyle{headings}
\mainmatter
\def\ECCVSubNumber{31}  

\title{Noise-Aware Merging of High Dynamic Range Image Stacks without Camera Calibration}
\titlerunning{Noise-Aware Merging of HDR Image Stacks without Camera Calibration}

\author{Param Hanji \and
Fangcheng Zhong \and
Rafa{\l} K. Mantiuk\index{Rafa{\l} Mantiuk}}
\authorrunning{P. Hanji et al.}
\institute{Department of Computer Science and Technology, University of Cambridge\\
\email{\{pmh64,fz261,rkm38\}@cam.ac.uk}}


\maketitle

\begin{abstract}

A near-optimal reconstruction of the radiance of a High Dynamic Range scene from an exposure stack can be obtained by modeling the camera noise distribution. The latent radiance is then estimated using Maximum Likelihood Estimation. But this requires a well-calibrated noise model of the camera, which is difficult to obtain in practice. We show that an unbiased estimation of comparable variance can be obtained with a simpler Poisson noise estimator, which does not require the knowledge of camera-specific noise parameters. We demonstrate this empirically for four different cameras, ranging from a smartphone camera to a full-frame mirrorless camera. Our experimental results are consistent for simulated as well as real images, and across different camera settings.

\end{abstract}

\section{Introduction}

The dynamic range of a scene may far exceed the range of light intensities that a standard digital sensor can capture. The conventional way of capturing all the information for such a \ac{HDR} scene is with a stack of images taken with different exposure times. These are later combined in post-processing as part of the digital pipeline \cite{mann1994beingundigital,tsin2001statistical,granados2010optimal,hasinoff2010noise,Aguerrebere2014,Gallo2016}. The probabilistic photon registration and electronic processing in the camera will result in some variation in the values recorded in each pixel, which manifests as noise in images. Any method attempting to accurately estimate the scene radiance from multiple images strives to increase the dynamic range while simultaneously reducing such noise. In this paper, we provide a comprehensive analysis of how noise in images affects the performance of several scene radiance estimators \cite{Debevec1997a,granados2010optimal,hasinoff2010noise}. This work is restricted to static and well-aligned images and we do not consider the problems of pixel alignment and deghosting \cite{karaduzovic2014expert,tursun2015state}.

It has been shown that, under the assumption of a normal distribution, \ac{MLE} provides near-optimal estimates of the true radiance values \cite{Aguerrebere2014}. However, it does not offer a closed-form solution and running non-linear solvers on large images is impractical. For this reason, \ac{MLE} is typically approximated with an iterative \ac{EM} algorithm \cite{granados2010optimal}. We show that such a solver does not always converge to the correct \ac{MLE} solution and thus, may introduce an error in estimation. Another limitation of \ac{MLE} is that it is highly sensitive to the correct calibration of noise parameters \cite{Aguerrebere2014}. Motivated by these observations, we derive a much simpler, analytical estimator based on the Poisson nature of photon noise that is independent of camera-specific noise parameters and can, therefore, be used with any camera without requiring prior knowledge of its noise characteristic.

Starting with a multi-source noise model \cite{hasinoff2010noise,aguerrebere2013study}, we describe a calibration procedure to determine camera-specific noise model parameters. For experimentation, we generated synthetic \ac{HDR} stacks using physically accurate simulations with the noise parameters of real cameras. We rely on such simulations to compare the empirical biases and standard deviations of different estimators for the scene radiance.

The main contributions of this paper are:
\begin{itemize}
    \item A simple yet practical camera noise model, fitted for several cameras with both large (full-frame) and small (smartphone) sensors. 
    \item A recommendation to use an estimator based on the Poisson nature of photon noise, which performs as well as near-optimal \ac{MLE} estimators for the usable dynamic range.
    \item An empirical validation showing that estimating the sensor noise characteristic is unnecessary when merging HDR images in a noise-aware manner.
    \item An extended analysis showing that the recommended estimator is robust to high camera noise and is a suitable choice for low-light \ac{HDR} photography.
\end{itemize}

\section{Related Work}
\label{sec:background}

Early \ac{HDR} reconstruction methods \cite{mann1994beingundigital,mitsunaga1999radiometric,robertson2003estimation,akyuz2007noise,Debevec1997a} focused on inverting the \ac{CRF}. This is because camera manufacturers did not historically provide access to unprocessed and uncompressed RAW images. Most estimators proposed were weighted averages of the linearized pixel values, where the weights were functions of the inverse and the derivative of the \ac{CRF}. Debevec and Malik proposed a hat-shaped function that assigns higher weights to linearized pixels near the middle of the intensity range \cite{Debevec1997a}. All these methods do not account for camera noise and therefore provide sub-optimal estimations of \ac{HDR} pixel values. We refer the reader to chapter three of the book \emph{\ac{HDR} Video} for a detailed discussion on \emph{Stack-Based Algorithms for HDR Capture and Reconstruction} \cite{Gallo2016}.

The first \ac{HDR} estimation method that used a noise model was proposed by Tsin et al. \cite{tsin2001statistical}. They proposed to combine images with weights equal to the ratios of the respective exposure times and standard deviations, measured directly from the images. In a later work, Granados et al. \cite{granados2010optimal} showed that Tsin et al.'s method was sub-optimal under a compound-normal noise assumption as pixels near the saturation point were given smaller than ideal weights despite having the highest \ac{SNR}. 
Debevec and Malik's hat-shaped weighting function \cite{Debevec1997a} also suffers from this limitation of under-weighting pixels close to the saturation point. 
Kirk and Anderson \cite{kirk2006noise} proposed an \ac{MLE} based weighting scheme using a simple noise model. This was later extended to a more complete model that incorporated noise from several sources by Hasinoff et al. \cite{hasinoff2010noise}. Granados et al. \cite{granados2010optimal} noted that the true \ac{MLE}-based estimator does not have an analytical solution and used the \ac{EM} algorithm for a more accurate estimation than other similar works.

Aguerrebere et al. \cite{Aguerrebere2014} compared the previously mentioned \ac{HDR} reconstruction methods and analyzed how far each of their variances were from the theoretical \ac{CRLB}. They concluded that the variance of MLE-based estimators, such as that of Granados et al. \cite{granados2010optimal}, were close to the \ac{CRLB} but the estimation could be easily affected by errors in noise parameter calibration. In this paper, we show that a comparable performance can be achieved by a simpler estimator without the need for camera calibration.

\section{Image Formation Pipeline}
\label{sec:imaging-pipeline}
We begin this section by explaining the capturing process, highlighting the probabilistic nature of camera noise. A more comprehensive description can be found elsewhere \cite{costantini2004virtual,aguerrebere2013study,Konnik2014}, but we include this overview for completeness. After introducing the sensor noise model, we describe the calibration procedure to estimate camera-specific noise model parameters.

\subsection{Sensor model}

\begin{wrapfigure}{r}{0.5\columnwidth}
    \centering
    \includegraphics[width=0.5\columnwidth]{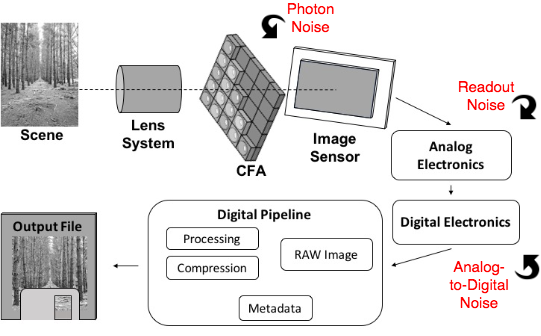}
    \caption{The illustration of an image formation pipeline that converts photons from a scene into images. Variance in pixel values arises from the noise added at different stages (marked in red).}
    \label{fig:sensor-noise}
\end{wrapfigure}

Photons from a scene are captured by the camera lens and pass through a \ac{CFA}, before being focused on an imaging sensor such as a \ac{CCD} or \ac{CMOS} sensor as depicted in Fig.~\ref{fig:sensor-noise}. When exposed for a fixed interval, the imaging sensor converts some incident photons to electrons. The number of electrons is proportional to the number of registered photons. These electrons accumulate to yield a voltage, which is processed by analog electronics. The next operation performed is amplification of the voltage based on the camera gain. The exposure time and gain are determined by the user-controllable shutter speed and ISO settings. Finally an \ac{ADC} digitizes the signal into discrete pixel intensities. Modern digital cameras provide access to this uncompressed, minimally processed data directly from the electronic imaging sensor in the form of RAW images.

\subsection{Noise model}
\label{subsec:noise-model}

RAW values are inaccurate measurements of the unknown scene radiance due to the potential saturation of the sensor and the addition of noise at various stages of the pipeline (Fig.~\ref{fig:sensor-noise}). To correctly reconstruct a scene in a noise-optimal manner, we model the probabilistic nature of noise in RAW images.

The process of photon registration by the sensor inherently follows a Poisson distribution \cite{Healey1994}. This leads to photon noise and is the first source of noise in our model. The other contributions to noise are signal-independent. Because the signal is amplified by some \emph{gain} before reaching the ADC, the signal-independent noise is typically split into pre-amplifier and post-amplifier components \cite{hasinoff2010noise}. Readout noise captures the voltage fluctuations while accumulating electrons and is amplified along with the signal. The last component, analog-to-digital noise, is added after amplification and is attributed to the quantization error. Digital sensors also exhibit fixed-pattern noise due to photo-response and dark-current non-uniformity \cite{aguerrebere2013study}. These sources of noise, however, are easy to compensate for as they are fixed for every sensor and are often removed by camera firmware from RAW images. We do not model fixed-pattern noise as it was not present in the images captured by our cameras. Other random sources of noise, such as temperature-dependent dark-current shot noise \cite{costantini2004virtual}, are accommodated in the signal-independent components. Moreover, previous works \cite{hasinoff2010noise,granados2010optimal} indicate that a simple, statistical noise model is sufficient for the problem of HDR radiance estimation.

Let the image be taken with an exposure time $t$ and gain $g$, and let $Y(p)$ be a random variable representing the final recorded value of the unknown scene radiance $\phi(p)$ at pixel $p$. The different sources of noise depicted in Fig.~\ref{fig:sensor-noise} motivate the decomposition of this random variable into a sum of three independent random variables. The first random variable is sampled from a Poisson distribution with a parameter equal to the number of incoming photons; the other two are sampled from zero-mean normal distributions. The first normally distributed component accounts for readout noise and has a standard deviation equal to $\sigma_\textrm{read}$ and the second component, parameterized by $\sigma_\textrm{adc}$, captures amplifier and quantization noise. Assuming that the pixel is not saturated,
\begin{equation}
    Y(p) \sim \textrm{Pois}(\phi(p)\,t)\,g\,k_c + \mathcal{N}(0,\sigma_\textrm{read})\,g\,k_c + \mathcal{N}(0,\sigma_\textrm{adc})\,k_c\,.
\label{eq:sample-physically-accurate}
\end{equation}
Each color channel has a different quantum efficiency for photon-to-electron conversion due to differences in the sensitivity of the sensor across the light spectrum. This is accounted for by the color coefficient $k_c$ where $c \in \{r,g,b\}$. Gain affects the Poisson random variable and the first normal random variable, while the $k_c$ is a multiplier on all three terms. The expected value and variance of $Y(p)$ can be written as:
\begin{equation}
\begin{aligned}
    E[Y(p)] & = \phi(p)\,t\,g\,k_c \\
    \textrm{var}(Y(p)) & = \phi(p)\,t\,g^2\,k_c^2 + \sigma_\textrm{read}^2\,g^2\,k_c^2 + \sigma_\textrm{adc}^2\,k_c^2 \\
    & = \underbrace{E[Y(p)]\,g\,k_c}_{\text{photon noise}} + \underbrace{\sigma_\textrm{read}^2\,g^2\,k_c^2 + \sigma_\textrm{adc}^2\,k_c^2\,}_{\text{static noise}}.
\end{aligned}
\label{eq:pixel-variance}
\end{equation}
Notice that the variance can be conveniently represented as a function of the expected value. We refer to the signal-independent component of the variance as static noise. Static noise is the same for all pixels of an image. When the radiance of the scene is close to zero, static noise can result in the underestimation of the true radiance, effectively making some pixel values negative. Camera manufacturers typically add an offset, called black-level, to ensure that RAW pixel values are positive. In all our experiments we subtract black level to operate on the actual measurements. 

\subsection{Noise parameter estimation}
\label{subsec:noise-parameters}
\begin{figure}[t]
    \begin{minipage}{0.35\columnwidth}
    \centering
    \includegraphics[width=\columnwidth]{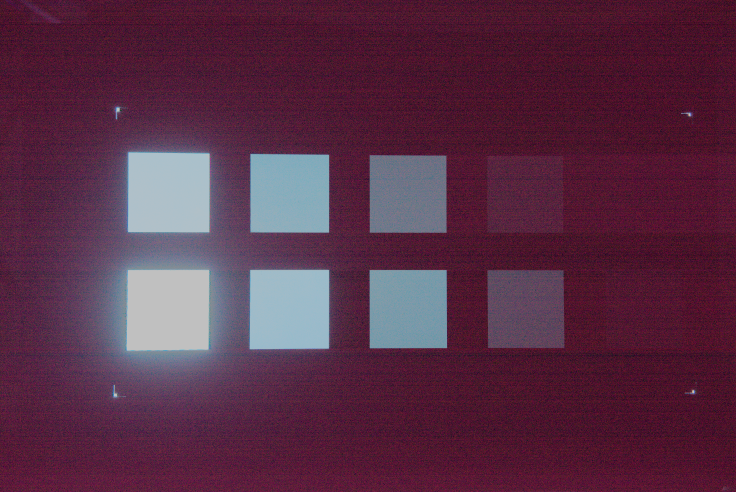} \\
    \includegraphics[width=\columnwidth]{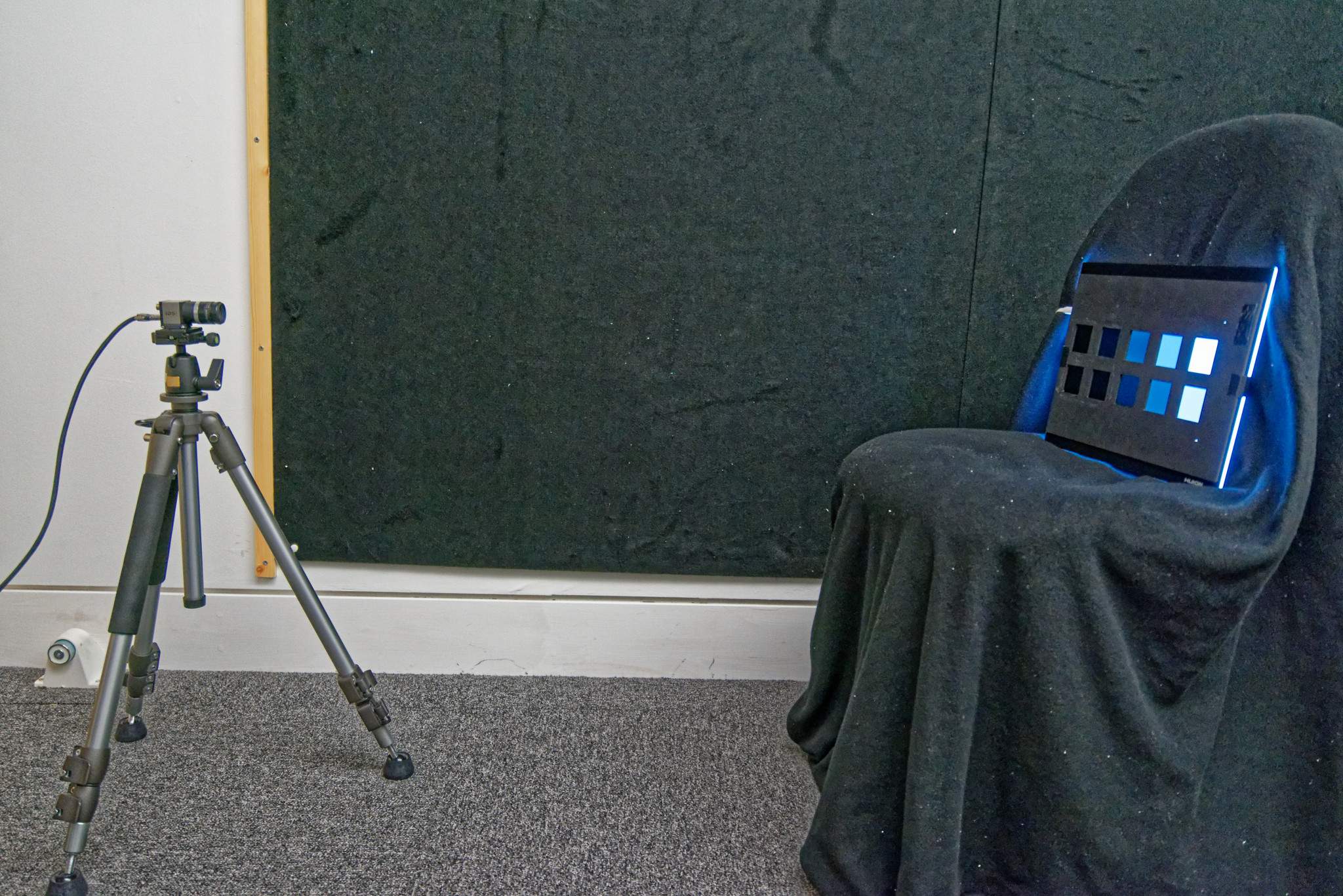}
    \end{minipage}
    \begin{minipage}{0.64\columnwidth}
    \centering
    \includegraphics[width=\columnwidth]{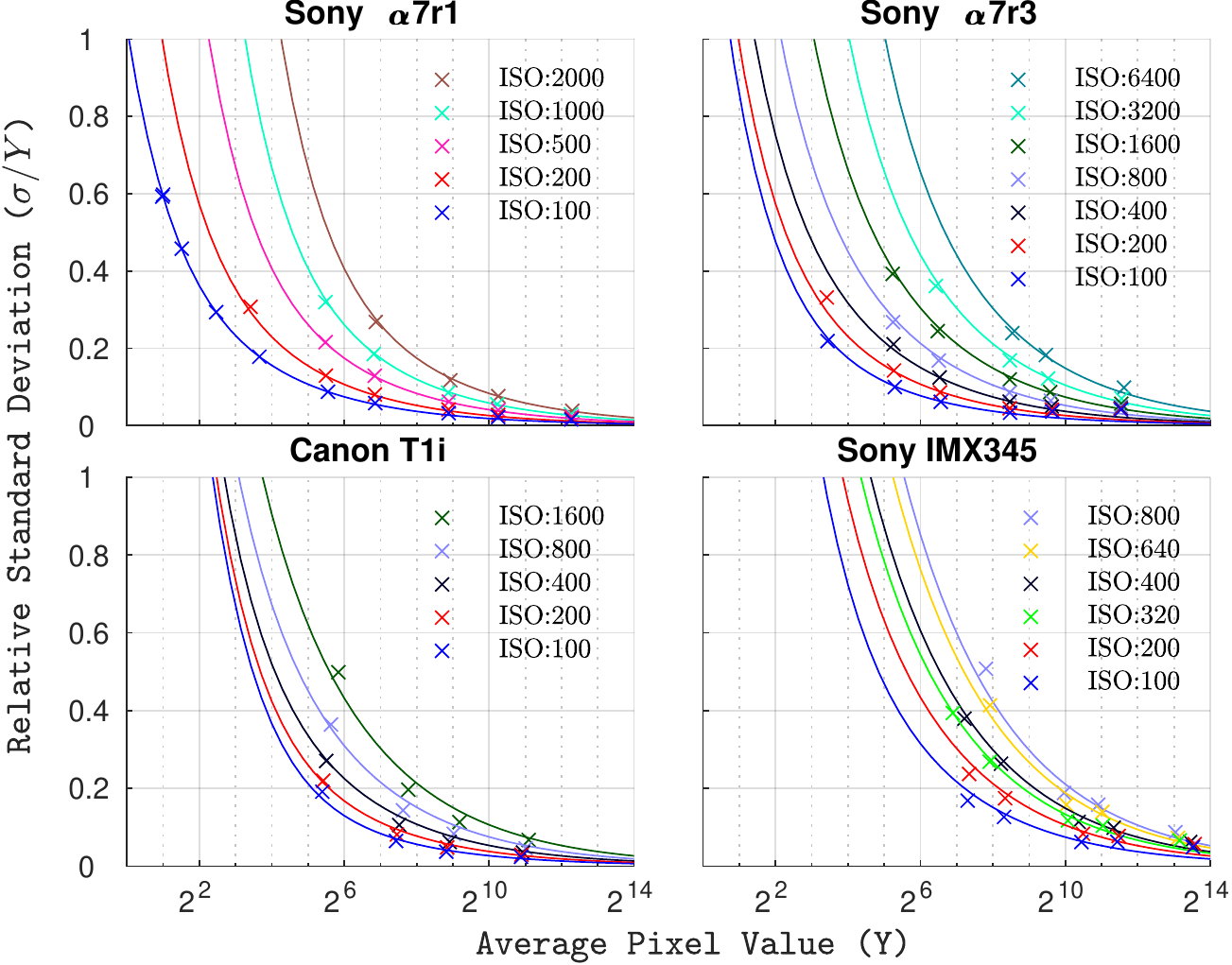}
    \end{minipage}
    \caption{The calibration target (top-left) and capture setup (bottom-left) used to measure the variance of sensor noise, and the corresponding fitted noise models for the green channel of different cameras (right). Relative standard deviation is plotted against the average pixel value recorded on a logarithmic scale. We control gain by changing ISO (different lines in each plot) since our noise equation (Eq.~\ref{eq:pixel-variance}) models how the variance of a pixel changes with gain. The \emph{crosses} represent measurements and the \emph{lines} are the model predictions. The RAW pixel values of all the cameras sensors have been scaled to $14$-bit values to enable the comparison of different sensors.}
    \label{fig:noise-plot}
\end{figure}

Let us consider Eq.~\ref{eq:pixel-variance}, the noise model, and how to estimate its camera specific parameters. Rather than measuring the noise added from various sources individually, we use a calibration target shown in the top-left image of Fig.~\ref{fig:noise-plot}. This is constructed by overlaying a uniform light source (a light box) with \ac{ND} filters of different transmittance values so that each square region emits a different radiance. The variance of pixels within each captured square provides an empirical measure of noise for a specific value of radiance.

We extracted several data points (RAW pixel mean and standard deviation pairs) and plotted them using crosses in Fig.~\ref{fig:noise-plot}. The mean of a large number of pixels contained within each square is used as a substitute for the expected value to fit Eq.~\ref{eq:pixel-variance}. To minimize the error in the expected value of the RAW pixel intensity of each square, we computed the average of all pixels within each square from a set of five images captured using the same settings. We captured several such sets starting with a base ISO of $100$ and an appropriate shutter speed such that none of the pixels were saturated. We assume that ISO 100 corresponds to the gain of 1. To capture subsequent sets of images, we doubled the ISO and halved the shutter speed every time to maintain the same mean intensity and to use the complete dynamic range of the camera sensor. The number of image sets varied from camera to camera and was typically between five and seven.

\begin{table}[t]
\caption{Noise parameters fitted for the tested cameras}
\centering
\begin{tabularx}{\columnwidth}{c c *{6}{Y}}
\toprule
\multirow{2}{*}{Camera}
 & \multirow{2}{*}{\thead{Sensor Size\\($mm$)}}
 & \multirow{2}{*}{\thead{Pitch\\($\mu m$)}}
 & \multicolumn{3}{c}{Color Coefficients}
 & \multirow{2}{*}{$\sigma_\textrm{read}$}
 & \multirow{2}{*}{$\sigma_\textrm{adc}$}\\
  & & & $k_r$ & $k_g$ & $k_b$\\
\midrule
Sony $\alpha$7r1 & 35.9 $\times$ 24 & 4.86 & 0.327 & 0.33 & 0.32 & 0.7 & 0.04 \\
Sony $\alpha$7r3 & 35.6 $\times$ 23.8 & 4.5 & 0.422 & 0.384 & 0.389 & 0.705 & 3.028 \\
Canon T1i & 22.3 $\times$ 14.9 & 4.69 & 1.363 & 1.183 & 1.153 & 0.928 & 5.005 \\
Sony IMX345 & 8.27 $\times$ 5.51 & 1.4 & 0.303 & 0.313 & 0.321 & 1.063 & 2.373 \\
\bottomrule
\end{tabularx}
\label{tab:camera_specifications}
\end{table}

We then used a nonlinear solver \cite{lagarias1998convergence} to estimate the 5 parameters of the noise model from Eq.~\ref{eq:pixel-variance}. Very noisy samples with \ac{SNR} less than one were excluded to ensure convergence. The fit for different cameras, shown in Fig.~\ref{fig:noise-plot}, demonstrates that the model can well explain the noise found in the tested cameras. Each plot shows the measured relative noise against recorded digital intensity, as well as fitted noise model for the green channel. Please refer to the supplementary material to view similar plots for the red and blue channels and also the individual contribution of each component of noise. For a better comparison, the digital values from each camera were rescaled such that the maximum pixel value of every image was set to $2^{14}-1$. This is the largest bit-rate registered among all our cameras. Estimated parameters for our calibrated cameras are given in Table~\ref{tab:camera_specifications}. Images with synthetic noise, generated using Eq.~\ref{eq:sample-physically-accurate} can be found in the supplementary material.

A few interesting observations can be made about the measured cameras. If we define the dynamic range as the ratio between the largest registered value and the smallest value whose \ac{SNR} is 1 (corresponding to $\sigma/Y=1$ in the plots in Fig.~\ref{fig:noise-plot}), we notice that the dynamic range differs substantially between the sensors. But it should be noted that these sensors differ in their pixel pitches and resolutions (see Table~\ref{tab:camera_specifications}). Hence, the effective amount of noise in images from the different sensors can vary even more when rescaled to the same resolution. The dynamic range of every sensor is also reduced with increased gain (ISO).

\subsection{Digital pipeline}
The RAW image from the sensor passes through the digital \ac{ISP} pipeline starting with black-level subtraction, followed by demosaicing, denoising, tone-mapping and compression \cite{Buckler2017}. Most of these stages perform nonlinear operations and alter the original readings significantly. Any attempt to recover the scene radiance should thus omit these digital operations. Working with RAW values is preferable as they are linearly related to the scene radiance. The only step needed is black-level subtraction to ensure this linear relationship between radiance and pixel values.

\section{HDR Radiance Estimation}
\label{sec:HDR-reconstruction}

In this section, we formulate the problem of estimating radiance values from a number of noisy sensor measurements. We use upper case letters to represent random variables and lower case letters to represent their observed values. Given a stack of RAW images $i=1,\,2,\,\dotsc,\,N$, let $y_i(p)$ represent the RAW value of pixel $p$ and image $i$ captured using a corresponding exposure time $t_i$ and gain $g_i$. To compensate for the differences in exposure time and gain, and to bring all exposures in the stack to the same scale, we represent their relative radiance as:
\begin{equation}
    X_i(p) = \frac{Y_i(p)}{t_i\,g_i\,k_c}\,,
\end{equation}
such that each $x_i(p)$ is an observation of the true value $\phi(p)$. The expected value of $X_i(p)$ is thus $\phi(p)$ and its variance is obtained by scaling Eq.~\ref{eq:pixel-variance}:
\begin{equation}
    \sigma^2_i(p) = \frac{\textrm{var}(Y_i(p))}{t_i^2\,g_i^2\,k_c^2} = \frac{\phi_i(p)}{t_i} + \frac{\sigma_\textrm{read}^2}{t_i^2} + \frac{\sigma_\textrm{adc}^2}{t_i^2\,g_i^2}\,.
    \label{eq:scaled-variance}
\end{equation}

All the estimators considered in this section assume that the images in the stack are perfectly aligned. This can be achieved by a global homography-based alignment \cite{tomaszewska2007image} or a local alignment based on optical flow \cite{liu2009beyond,zimmer2011optic,anderson2016jump}. 

\subsubsection{Uniform estimator}
The simplest estimator is the arithmetic mean of all the available samples and is referred to as the \emph{Uniform} estimator. This is obviously a poor estimator as pixel values from different images in the stack are sampled from different distributions and have different \ac{SNR}s.

\subsubsection{Hat-shaped estimator} The widely-used weighting scheme proposed by Debevec and Malik \cite{Debevec1997a} assigns higher weights to image pixel values in the middle of the intensity range. The weights are functions of tone-mapped pixel values that pass through the whole pipeline. Let the smallest and largest intensities that can be recorded by a particular sensor be $y_\textrm{min}$ and $y_\textrm{max}$ respectively. Approximating the \ac{CRF} by the gamma function, the weights are equal to
\begin{equation}
    w_i(p) = \begin{cases}
        y_i(p)^{1/\gamma} - y_\textrm{min}^{1/\gamma} + \epsilon, & y_i(p)^{1/\gamma} \leq {\displaystyle\frac{y_\textrm{min}^{1/\gamma} + y_{\textrm{max}\vphantom{i}}^{1/\gamma}}{2}}\\
        y_\textrm{max}^{1/\gamma} - y_i(p)^{1/\gamma} + \epsilon, & y_i(p)^{1/\gamma} > {\displaystyle\frac{y_\textrm{min}^{1/\gamma} + y_{\textrm{max}\vphantom{i}}^{1/\gamma}}{2}}\,,
        \end{cases}
\end{equation}
where $\epsilon=10^{-10}$ is a small constant added to ensure that the weights are strictly positive and prevent division by zero. $\gamma$ is generally set to 2.2. Pixels with values close to the noise floor or the saturation point are assigned lower weights. The weights are applied to the linearized pixel values to obtain the \emph{Hat-shaped} estimator:
\begin{equation}
    \hat{\phi}_\textrm{hat}(p) = \frac{\sum_{i=1}^N w_i(p)\,x_i(p)}{\sum_{i=1}^N w_i(p)}\,.
\end{equation}

\subsection{Maximum Likelihood Estimation}

Given the probabilistic image formation model described by Eq.~\ref{eq:sample-physically-accurate}, the best estimators are based on \ac{MLE}, which has been shown to be near-optimal for this problem \cite{Aguerrebere2014}. The problem is that the \ac{PDF} of each $X_i(p)$ is a convolution of the three independent density or mass functions, and does not have a close-form expression.

\subsubsection{Variance-weighted estimator} The noise-based estimator introduced in \cite{kirk2006noise} and extended in \cite{hasinoff2010noise} assumes that the Poisson component of each $X_i(p)$ can be approximated by a normal distribution. Eq.~\ref{eq:sample-physically-accurate} then simplifies to the sum of three normally distributed random variables, which is also normally distributed. The log-likelihood function to be maximized simplifies to:
\begin{equation}
    \ln{\mathcal{L}_\mathcal{N}(\phi(p))} = \sum_{i=1}^N \ln{\frac{1}{\sqrt{2\,\pi\,\sigma^2_i(p)}}} + \sum_{i=1}^{N} \frac{-(x_i(p) - \phi(p))^2}{2\,\sigma^2_i(p)}\,.
    \label{eq:normal-likelihood}
\end{equation}
where $\sigma^2_i(p)$ is the variance of $X_i(p)$ from Eq.~\ref{eq:scaled-variance}. When $\sigma^2_i(p)$ and $\phi(p)$ are independent, the \ac{MLE} has a simple form: 
\begin{equation}
    \hat{\phi}_\textrm{var}(p) = \frac{\sum_{i=1}^N \frac{x_i(p)}{\sigma^2_i(p)}}{\sum_{i=1}^N \frac{1}{\sigma^2_i(p)}}\,.
    \label{eq:est-weighted-average}
\end{equation}

The problem is that $\sigma^2_i(p)$ and $\phi(p)$ are not independent as $\sigma^2_i(p)$ is a function of $\phi(p)$ due to photon noise (refer to Eq.~\ref{eq:scaled-variance}). This requires another simplifying assumption, that $\sigma^2_i(p)$ can be estimated using a single observation $y_i(p)$. However, when $E[Y_i(p)]$ is approximated by $y_i(p)$, the variance computed using Eq.~\ref{eq:scaled-variance}, can be zero or negative. This may happen because $y_i(p)$ can be negative and larger in magnitude than the static noise. In our implementation, the weights $\frac{1}{\sigma^2_i(p)}$ are replaced in such instances with a small value of $\epsilon = 10^{-10}$.

\subsubsection{Iterative Expectation Maximization (EM)}
A simple method that produces a computationally efficient \ac{MLE} solution for Eq.~\ref{eq:normal-likelihood} is the \ac{EM} algorithm. In this iterative approach, proposed by Granados et al. \cite{granados2010optimal}, $\phi(p)$ is initialized as the mean of $y_i(p)$. Then variances for every exposure are calculated according to Eq.~\ref{eq:scaled-variance} and a better estimate of $\phi(p)$ is found using the new variances according to Eq.~\ref{eq:est-weighted-average}. The alternating procedure is repeated until converge.

\subsubsection{Full MLE}
\label{subsec:complete}
The \ac{MLE} for Eq.~\ref{eq:normal-likelihood} can be estimated without making any additional assumptions using a non-linear optimization method. Although this is the most accurate estimator, running such a solver for each pixel is too computationally expensive to be used in practice. We include this estimator to show how close other estimator are to the true \ac{MLE} solution.

\subsubsection{Normal Photon Noise Estimator (NPNE)} Eq.~\ref{eq:normal-likelihood} has an analytical solution if we assume that the variance is only due to photon noise and there is no static noise. Such a simplification is justified because the overall noise is dominated by the photon noise component for the usable range of $\phi(p)$. Setting static noise to zero and maximizing Eq.~\ref{eq:normal-likelihood} yields the estimator:
\begin{equation}
    \hat{\phi}_\textrm{npne}(p) = \frac{\sqrt{\sum_{i=1}^N x^2_i(p)\,t_i \cdot \sum_{i=1}^N t_i + N^2} - N}{\sum_{i=1}^N t_i}\,.
    \label{eq:est-mle-normal}
\end{equation}

\subsubsection{Poisson Photon Noise Estimator (PPNE)}
Setting static noise to zero also enables us to simplify the \ac{PDF} of each $X_i(p)$ and derive an estimator that maximizes the likelihood without the normal approximation to the Poisson distribution. This means that the random variables are sampled from:
\begin{equation}
    X_i(p) \sim \frac{\textrm{Pois}(\phi(p)\,t_i)}{t_i}\,.
\end{equation}
And the new log-likelihood function to be maximized is:
\begin{equation}
    \ln{\mathcal{L}_\textrm{Pois}(\phi(p))} = \sum_{i=1}^{N} x_i\,t_i\,\ln{\phi\,t_i}\, - \sum_{i=1}^{N} \phi\,t_i - \sum_{i=1}^{N} \ln{(x_i\,t_i)!}\,.
\end{equation}
The last sum does not depend on $\phi(p)$ and can safely be ignored. The resulting Poisson Estimator takes on the simple form:
\begin{equation}
\begin{aligned}
    \hat{\phi}_\textrm{ppne}(p) 
    &= \frac{\sum_{i=1}^N x_i(p)\,t_i}{\sum_{i=1}^N t_i}
\end{aligned}
\label{eq:est-mle-poisson}
\end{equation}

Such an estimator is a classical choice in the imaging industry \cite{Aguerrebere2014}. Here, we demonstrated how it can be derived from the assumption of Poisson noise. Additionally, in the supplementary material, we employ the \textit{Lehmann-Scheffe theorem} to show that \emph{PPNE} is the unique \ac{MVUE} under the assumption of zero static noise. This is an important result, since estimators produced by \ac{MLE} are generally not guaranteed to be unbiased or to have the minimum variance.

\subsubsection{Estimators and noise parameters}

An advantage of \emph{NPNE} and \emph{PPNE} estimators that ignore static noise is that they do not require knowledge of the noise parameters to provide accurate estimates. Eqs.~\ref{eq:est-mle-normal} and~\ref{eq:est-mle-poisson} are functions of $x_i(p)$, which depends only on the parameter $k_c$. And, since $k_c$ is the same for all exposures, it is effectively a constant multiplier, which does not affect relative radiance values.

\section{Comparison of Estimators --- Simulation}
\label{subsec:comparison-estimators}

We empirically compare the calibration-independent estimators, \emph{NPNE} and \emph{PPNE}, to the classical \emph{Uniform} and \emph{Hat-shaped} estimators \cite{Debevec1997a} and state-of-the-art MLE-based estimators, \emph{Variance-weighted} \cite{hasinoff2010noise} and \emph{Iterative EM} \cite{granados2010optimal,Aguerrebere2014}. We rely on \ac{MC} methods and simulate 10,000 \ac{HDR} exposure stacks. Each stack consists of three exposures separated by five stops in exposure time and a constant gain of 8. Starting with 100 logarithmically-spaced ground truth values of radiance, spanning a dynamic range of 24 stops, we simulated the probabilistic camera capture process described in Section~\ref{subsec:noise-model}. Our noisy samples were generated using the parameters of calibrated cameras from Table~\ref{tab:camera_specifications}.

\begin{figure}[t]
    \centering
    \includegraphics[width=\columnwidth]{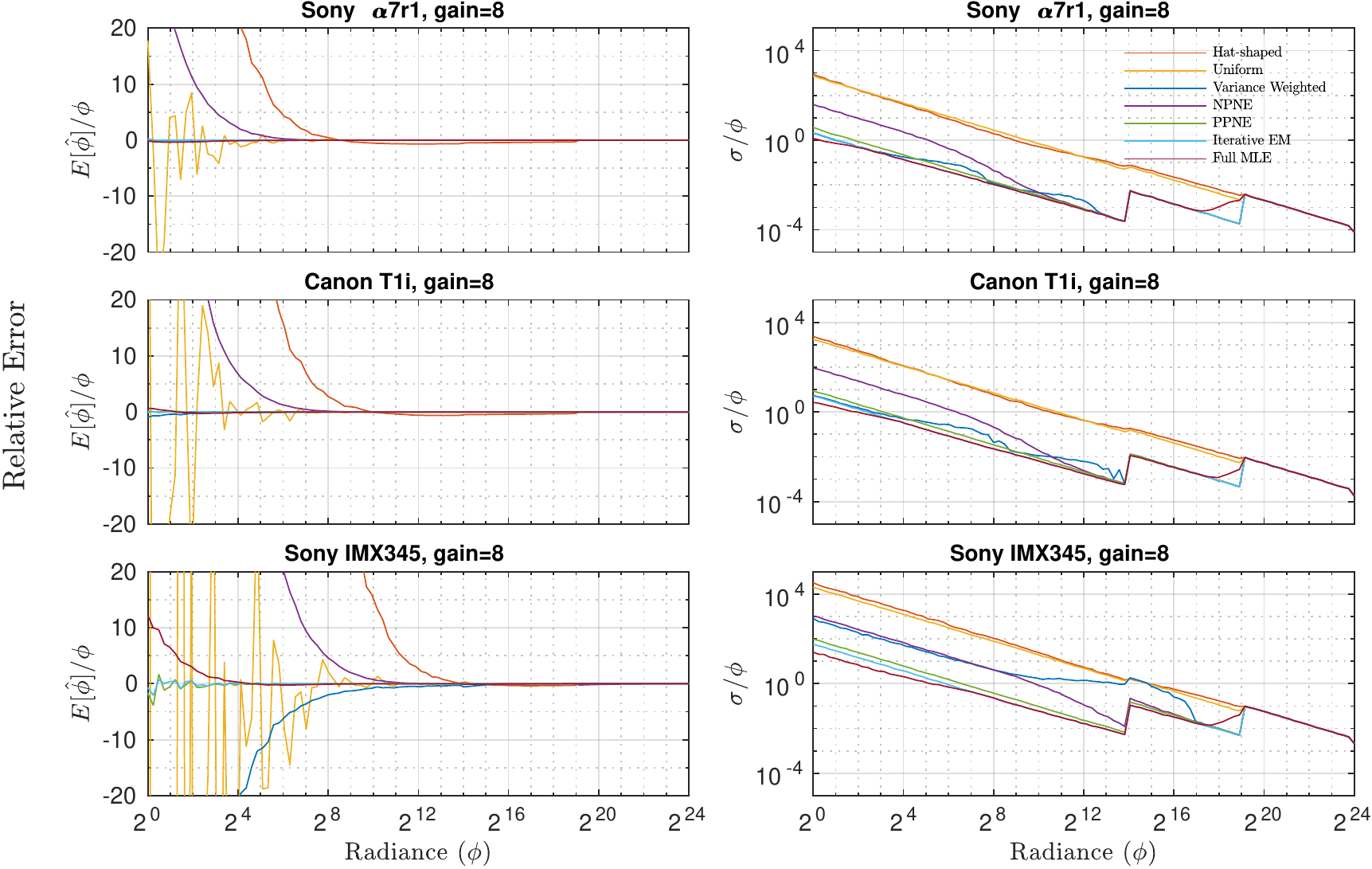}
    \caption{Monte Carlo simulation results: relative errors of the different estimators (\emph{color lines}) for logarithmically spaced values of the ground truth scene radiance, $\phi$ (\emph{x-axis}). The errors arise due to non-zero biases (left column of plots) and non-zero standard deviations (right column of plots). The simulation was performed for three exposures, spaced five stops apart.}
    \label{fig:estimator-error-1d}
\end{figure}

Fig.~\ref{fig:estimator-error-1d} shows relative errors of the compared estimators for three calibrated cameras. We show relative quantities because they better correspond with perceived magnitudes (Weber's law). Notice that when the radiance $\phi$ is large, all the estimators are unbiased, and the error is solely due to the relative standard deviation. As expected, both the bias and standard deviation are highest for the smartphone sensor (Sony IMX345), and smallest for the full-frame mirrorless camera (Sony $\alpha$7r3). The sawtooth patterns visible in the standard deviation plots are due to the three exposures: the highest five stops are captured only in images with the shortest exposure time and, therefore, the error drops at $\phi \approx 2^{19}$ and $\phi \approx 2^{14}$ when data from the second and third images become available.

The results in Fig.~\ref{fig:estimator-error-1d} confirm that the estimators that do not account for noise (\emph{Uniform} and \emph{Hat-shaped}) result in the highest bias and the highest amount of noise. The bias of the \emph{Uniform} estimator appears unstable at low radiance values because the estimate is dominated by noise. The popular \emph{Variance-weighted} estimator performs reasonably well for the high quality sensor (Sony $\alpha$7r3) but results in a noisy estimate with a negative bias for the two other sensors. The negative bias is due to clamping of weights since variances can not be negative. As expected, the \emph{Full MLE} achieves the best performance. It is, however, much more computationally expensive. Notice that, the \emph{Iterative EM} estimator does not converge to the same solution as the \emph{Full MLE} for lower radiance values. 

The most interesting results are seen for the two estimators that account only for photon noise. The estimator that assumes photon noise is normally distributed (\emph{NPNE}) introduces a positive bias and results in a higher standard deviation than the estimator that assumes Poisson photon noise distribution (\emph{PPNE}). Overall, the error of the \emph{PPNE} is comparable with that of the \emph{Iterative EM} estimator, which is the best estimator used in practical applications. The relative standard deviation of noise is only marginally higher than that of the \emph{Iterative EM} but only for very low pixel values. However, it has a major advantage over the \emph{Iterative EM} as it does not require knowledge of the noise parameters. Therefore, it can be used with any camera without prior calibration, as long as the camera noise characteristic can be well explained by our model (Eq.~\ref{eq:sample-physically-accurate}).

\begin{figure}[!t]
    \centering
    \includegraphics[width=\columnwidth]{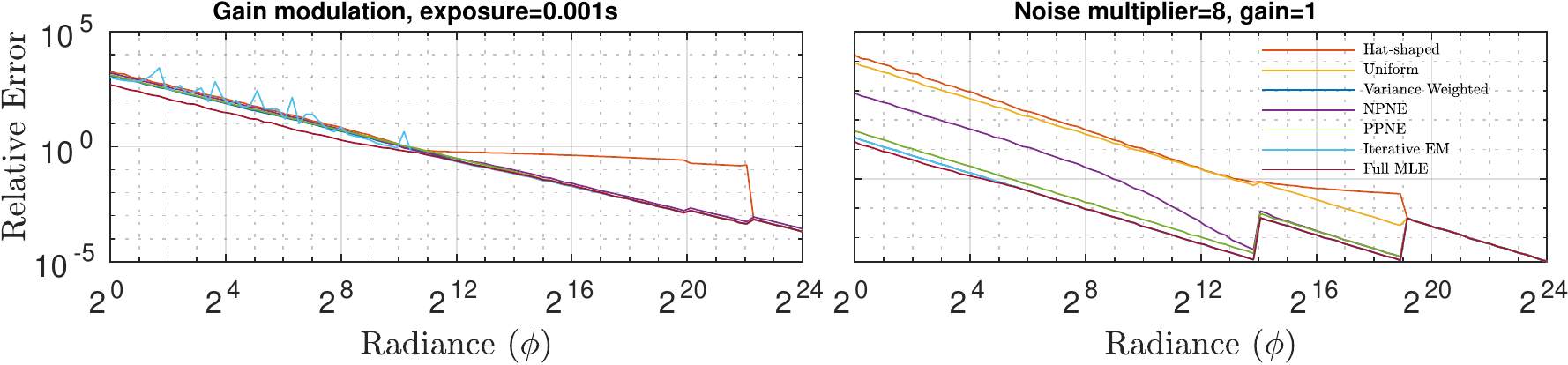}
    \caption{Relative errors of estimators for two additional scenarios for the Sony $\alpha$7r3: using a stack of images with different gain and same exposure time (left) and the effect of increasing the static noise by $8\times$ (right). The increased error of the \emph{Hat-shaped} estimator is due to a small negative bias (see left column of Fig.~\ref{fig:estimator-error-1d}).}
    \label{fig:iso-and-noise}
\end{figure}

\subsection{Gain modulation}
\label{subsec:gain-mod}

An alternate acquisition strategy for capturing \ac{HDR} scenes is modulating gain while maintaining the same exposure time and consistent motion blur across images \cite{hajisharif2015adaptive}. A detailed analysis of the increase in dynamic range and \ac{SNR} with the number of images captured for different strategies is presented in the supplementary material. Here, we compare the estimators on a stack of gain modulated captures. The plots on the left of Fig.~\ref{fig:iso-and-noise} indicate that the performance of most estimators is very similar, except for the better performance of \emph{Full MLE} for low radiance and failure of the \emph{Hat-shaped} estimator for medium radiance values. 

\subsection{Robustness to noise}
\label{subsec:high-noise}

We validated the performance of the estimators for exposure stacks captured with increased camera noise. The noise in input image stacks was artificially increased by amplifying the contribution of static noise $2\times, 4\times$ and $8\times$ the measured value for the Sony $\alpha$7r3 sensor. See the right side of Fig.~\ref{fig:iso-and-noise} for the simulation with $8\times$ static noise and the supplementary for other multipliers. These additional \ac{MC} simulations, confirm that the relative performance of \emph{PPNE} does not noticeably degrade with noise. Its relative error is very similar to the calibration-sensitive \emph{Iterative EM} estimator even when the input is very noisy. Since the proportion of static noise is much greater in low-luminance conditions, these results indicate that \emph{PPNE} is suitable for low-light \ac{HDR} photography.

\section{Comparison of Estimators --- Real Images}

\begin{figure}[t]
    \centering
    \includegraphics[width=\columnwidth]{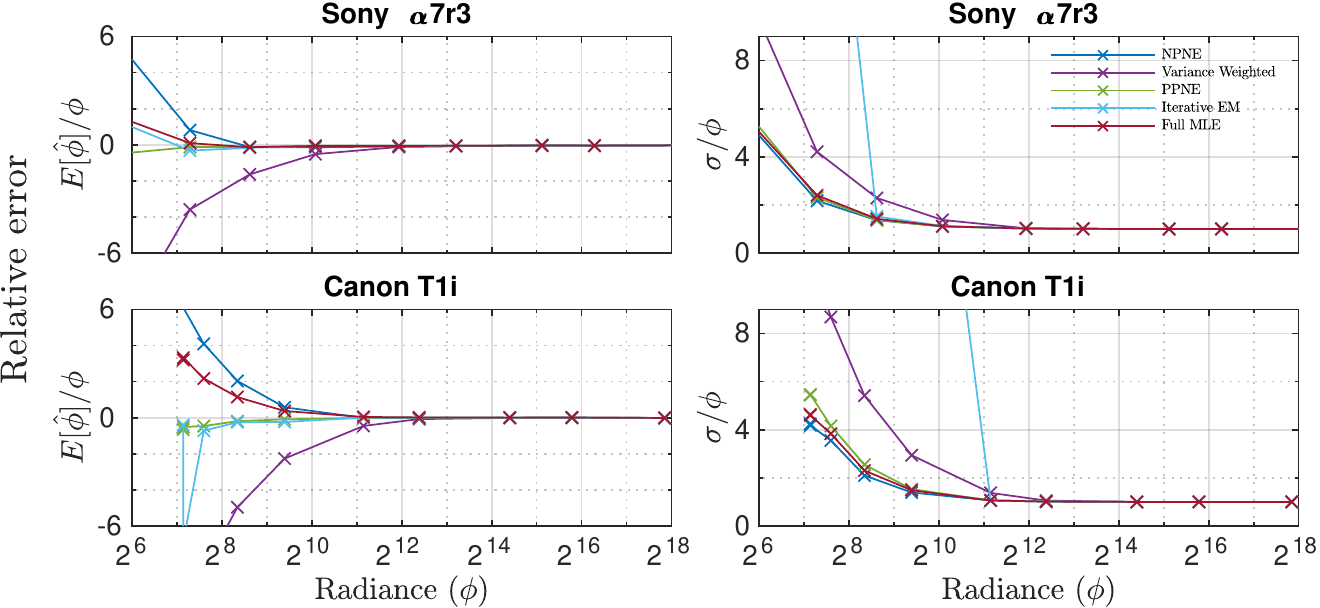}
    \caption{Results for real images: observed relative biases and standard deviations for the green channel of two cameras. The sudden increase in standard deviation at $\phi \approx 2^{11}$ for the \emph{Iterative EM} estimator in the top-right plot is due to the non-convergence of the \ac{EM} algorithm for low inputs.
    }
    \label{fig:estimator-error-real}
\end{figure}

To make sure that our simulation results are not the outcome of wrong assumptions about camera noise, we measured the errors of different estimators on real data using a stack of images of our calibration target (see Fig.~\ref{fig:noise-plot}-left). The stack is composed of images captured with different exposure times and gains as described in Section~\ref{subsec:noise-parameters}. The reference radiance $\phi$ was calculated as the average of all pixels in each square of the target. Fig.~\ref{fig:estimator-error-real} shows the error, due to bias and standard deviation, of the \ac{HDR} estimations for our calibrated cameras. Here we see a similar pattern as in Fig.~\ref{fig:estimator-error-1d}, where the standard deviation of the \emph{Variance-weighted} estimator is much higher than other MLE-based estimators and it has a large negative bias at low radiance. The performance of the analytical \emph{PPNE} is very similar to that of the \emph{Iterative EM} and the \emph{Full MLE}.

\begin{figure}[t]
    \centering
    \includegraphics[width=\columnwidth]{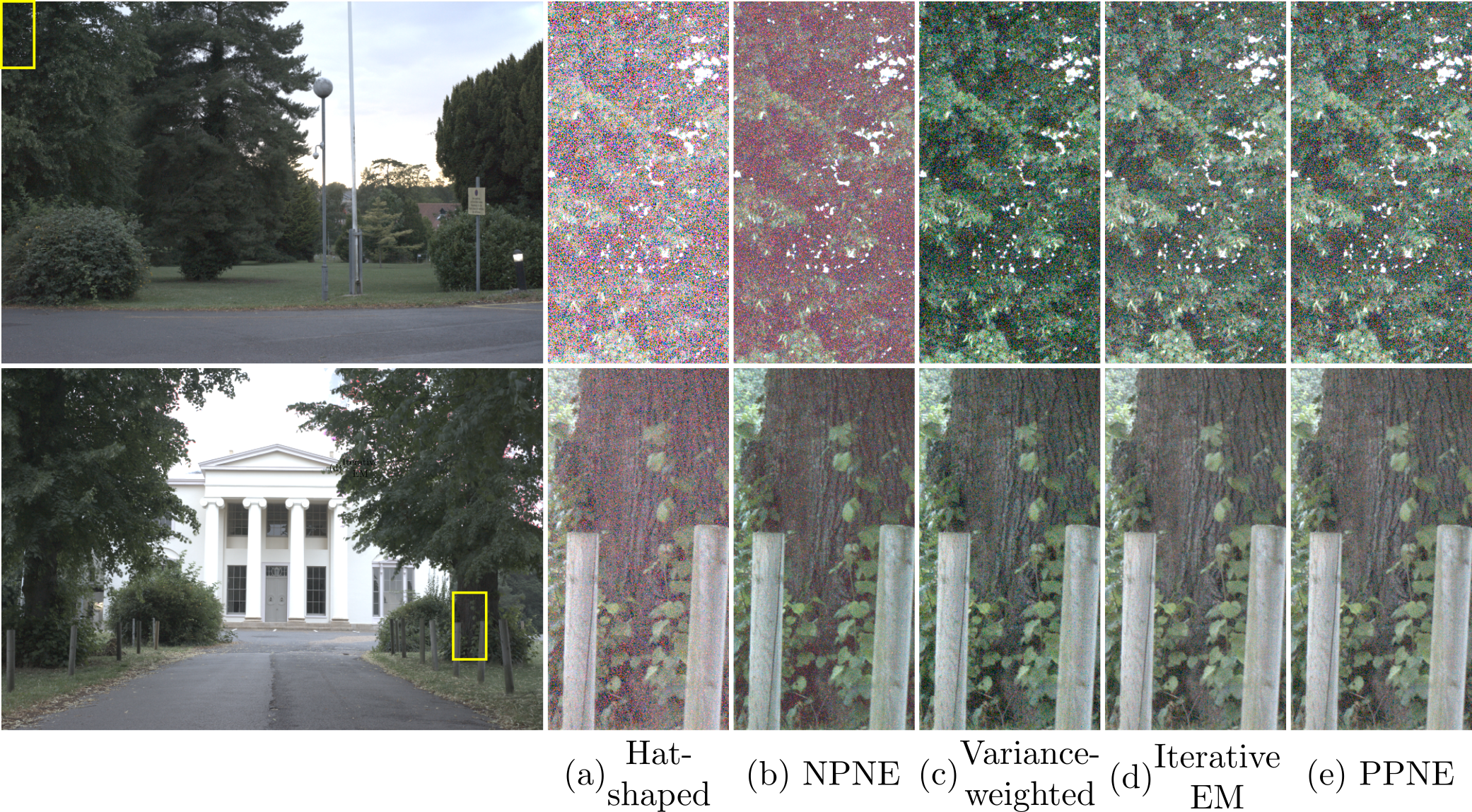}
    \caption{\ac{HDR} reconstructions of outdoor scenes, ``Trees" and ``House", using different estimators given exposure stacks of three images captured by the Sony $\alpha$7r3 at ISO 6400 and gamma-encoded for visualization ($\gamma=2.2$). The positive bias of the \emph{Hat-shaped} estimator and \emph{NPNE} as well as the negative bias of the \emph{Variance-weighted} estimator are visible in the dark regions. The images produced by the iterative EM estimator and \emph{PPNE} are almost identical. Refer to the supplementary for the ``Cottage" and ``Street" scenes. In all the scenes, the shortest exposure times were deliberately set to a small value to produce noisy images and test the robustness of the estimators.}
    \label{fig:primitives-and-colorchecker}
\end{figure}

\subsection{Qualitative and quantitative comparison on complex images}
Next, we show visual differences in HDR images due to the choice of the estimator in challenging conditions. We captured several scenes with three exposure times, spaced two stops apart. The images are processed with different estimators and show substantial differences, as depicted in Fig.~\ref{fig:primitives-and-colorchecker}. The \emph{Hat-shaped} weights resulted in a noisy image with a positive bias (Fig.~\ref{fig:primitives-and-colorchecker}a) in dark regions. \emph{NPNE} substantially reduced the amount of noise but still produced the bias (Fig.~\ref{fig:primitives-and-colorchecker}b). The \emph{Variance-weighted} estimator produced a noisier image than \emph{NPNE} (Fig.~\ref{fig:primitives-and-colorchecker}c) and also introduced a negative bias. This made some pixel values darker than they should be, resulting in an accidental increase in contrast. Merged images of \emph{PPNE} (Fig.~\ref{fig:primitives-and-colorchecker}d) and the \emph{EM} estimator (Fig.~\ref{fig:primitives-and-colorchecker}e) show the least amount of noise and smallest bias. For additional indoor and outdoor scenes, please refer to the supplementary material. Apart from the independence to noise-parameters, another advantage over the \emph{Iterative EM} estimator is the reduced computation time of \emph{PPNE} due to its analytical form. This does not make much of a difference for the simple logarithmic gradient, but is important for high-resolution images captured by a DSLR.

\begin{table}[t]
\resizebox{\textwidth}{!}{
\begin{tabular}{c|cccc|cccc|cccc}
\toprule
\multirow{2}{*}{Estimator} & \multicolumn{4}{c|}{PU-PSNR}      & \multicolumn{4}{c|}{PU-SSIM}       & \multicolumn{4}{c}{HDR-VDP-3}        \\
\cline{2-13}
                           & House  & Trees  & Cottage & Street & House & Trees & Cottage & Street & House & Trees & Cottage & Street \\
\hline
Hat-shaped                 & 20.266 & 13.241 & 18.516 & 9.019      & 0.636 & 0.489 & 0.51 & 0.302     & 5.719 & 5.39 & 5.013 & 4.638  \\
Var-weighted               & 16.412 & \third{18.191} & \second{22.476} & 13.267      & 0.624 & 0.497 & \third{0.578} & \second{0.512}     & \third{6.049} & 5.946 & \third{5.846} & \third{5.581}  \\
NPNE                       & \third{24.326} & 17.451 & 13.588 & \third{13.331}  & \second{0.681} & \first{0.543} & 0.374 & 0.289     & 6.027 & \third{6.114} & 5.391 & 5.304  \\
PPNE                       & \second{24.428} & \first{18.876} & \third{22.336} & \second{13.361}      & \first{0.72} & \second{0.538} & \second{0.607} & \third{0.432}     & \second{6.255} & \second{6.253} & \second{5.904} & \second{5.802}  \\
EM                         & \first{24.922} & \second{18.339} & \first{23.059} & \first{13.962}      & \first{0.72} & \third{0.52} & \first{0.614} & \first{0.524}     & \first{6.273} & \first{6.37} & \first{5.983} & \first{5.928} \\
\bottomrule
\end{tabular}
}
\caption{The reconstruction error for images in Fig.~\ref{fig:primitives-and-colorchecker} and Fig.~4 in the supplementary. The error is computed using HDR image quality metrics: PU-PSNR, PU-SSIM \cite{aydin_hvei08} and HDR-VDP-3 \cite{mantiuk2011hdr} (v3.0.6, Q-values). For all the metrics, a higher value denotes higher quality. In each column, the highest value has a gold background, the second-best has a silver background and the third has a bronze background. Overall, \emph{PPNE} is the second-best estimator and it is narrowly outperformed by the calibration-sensitive \emph{EM} estimator.}
\label{tab:iqa}
\end{table}

Finally, we report quality scores for three HDR image quality metrics in Table~\ref{tab:iqa}. The test images were obtained from three exposures, captured at high ISO setting and merged with each estimator while the reference images were obtained by merging five exposures with the \emph{EM} estimator (the most accurate). The results confirm the findings of other experiments; \emph{PPNE} produces results that are only marginally worse than those of \emph{EM}, even though \emph{EM} was used to generate the reference images.

\section{Conclusions}

Although the state-of-the-art HDR reconstruction methods advocate using \ac{MLE} solvers that require accurate camera parameters, we demonstrate that they provide little advantage over the simple Poisson noise estimator, which does not require camera noise calibration. We show that the Poisson noise estimator is unbiased and its standard deviation is only marginally higher that of the near-optimal \ac{MLE} solution for very low pixel values. Such a difference is unlikely to be noticed in complex images. For a simplified noise model, the Poisson estimator is provably \ac{MVUE}. Furthermore, we show how each estimator can be derived making different simplifying assumptions about the camera noise model, and we illustrate the relative errors of the estimators using gain modulation and under increased static noise. In all our experiments, the Poisson noise estimator was consistently among the best performing estimators.

\section*{Acknowledgement}
We would like to thank Minjung Kim and Maryam Azimi for their advice on the paper. This project has received funding from the European Research Council (ERC) under the European Union's Horizon 2020 research and innovation programme (grant agreement N$^\circ$ 725253--EyeCode).

\bibliographystyle{splncs04}
\bibliography{references}

\end{document}